\documentclass[journal]{IEEEtran}

\usepackage{algorithmic, graphicx, amsfonts}
\usepackage{amsmath, amssymb}
\usepackage{cite, array, multirow}
\usepackage{siunitx}
\usepackage{xcolor, color, hhline, stfloats}
\usepackage{lipsum}
\usepackage{gensymb}
\usepackage{verbatim}
\ifCLASSINFOpdf
\else
\fi

\begin{document}
\title{Temporal-Spatial Neural Filter: Direction Informed End-to-End Multi-channel Target Speech Separation}

\author{Rongzhi~Gu 
        and Yuexian~Zou \\
        Peking University Shenzhen Graduate School, Shenzhen, China
}

\markboth{}%
{}

\maketitle

\begin{abstract}

Target speech separation refers to extracting the target speaker's speech from mixed signals.
Despite the recent advances in deep learning based close-talk speech separation, the applications to real-world are still an open issue. Two main challenges are the complex acoustic environment and the real-time processing requirement.
To address these challenges, we propose a temporal-spatial neural filter, which directly estimates the target speech waveform from multi-speaker mixture in reverberant environments, assisted with directional information of the speaker(s).
Firstly, against variations brought by complex environment, the key idea is to increase the acoustic representation completeness through the jointly modeling of temporal, spectral and spatial discriminability between the target and interference source.
Specifically, temporal, spectral, spatial along with the designed directional features are integrated to create a joint acoustic representation.
Secondly, to reduce the latency, we design a fully-convolutional autoencoder framework, which is purely end-to-end and single-pass. All the feature computation is implemented by the network layers and operations to speed up the separation procedure.
Evaluation is conducted on simulated reverberant dataset WSJ0-2mix and WSJ0-3mix under speaker-independent scenario. Experimental results demonstrate that the proposed method outperforms state-of-the-art deep learning based multi-channel approaches with fewer parameters and faster processing speed. Furthermore, the proposed temporal-spatial neural filter can handle mixtures with varying and unknown number of speakers and exhibits persistent performance even when existing a direction estimation error. Codes and models will be released soon. 
\end{abstract}


\IEEEpeerreviewmaketitle

\section{Introduction}
\IEEEPARstart{S}{peech-driven}
human-machine interaction has progressed up to real-world scenarios today, with applications to smart speakers, intelligent conference room and virtual assistant, etc. Speech separation has become a major concern as the front-end for robust speech interaction.

Leveraging the advances in deep learning, close-talk monaural speech separation has achieved a great progress in recent years \cite{hershey2016deep, yu2017permutation, chen2017deep,wang2018alternative,luo2018tasnet, luo2019convtasnet}. Most existing approaches formulate speech separation problem in time-frequency (T-F) domain, owing to T-F sparsity and auditory masking effect \cite{colle1976acoustic, pfander2010sparsity}. Specifically, the network is trained to learn a mapping from mixture features to a T-F mask, where each element indicates the dominance of a source at each T-F bin of the mixture spectrogram \cite{wang2018supervised}. 
Due to the complexity of phase reconstruction, most methods decouple the magnitude and phase part. The mask is estimated only based on the magnitude spectra and employed to reconstruct the speech waveform along with the mixture phase. However, many researches show that there is still nonnegligible error even reconstructing speech with ideal magnitude and mixture phase \cite{le2019phasebook, takahashi2018phasenet, choi2019phase}. This introduces a performance upper bound for methods based on T-F magnitude masking. Although the significance of phase retrieval has been addressed by many recent works and achieved improved performance \cite{le2019phasebook, takahashi2018phasenet, choi2019phase, wang2018end, wichern2018phase}, the estimated phase remains suboptimal on speech separation task. Also, complex phase reconstruction algorithms may increase the system latency.

To avoid the phase problem and speed up the separation process, an increasing number of researchers focus on time-domain end-to-end speech separation \cite{luo2018tasnet, luo2019convtasnet, shi2019furcax}. One of the representative work is fully convolutional time-domain audio separation network (Conv-TasNet) \cite{luo2019convtasnet}. Conv-TasNet formulates the separation task in a real-valued high-dimensional space, which is assumed to contain both magnitude and phase information. Specifically, Conv-Tasnet employs a linear encoder to map the mixture chunk to a high-dimensional representation that optimized for speech separation. Separation is achieved by estimating the weight (mask) of each speaker in the high-dimensional space. See Section \ref{sec:single-channel_tss} for details.

Despite the significant improvement achieved by these end-to-end close-talk speech separation methods, there remains two main problems towards real-world applications: \emph{i}) The close-talk multi-speaker mixture model originates from the linear instantaneous mixture model, where the observed mixtures are linear combinations of the sources. While at far-field, the general mixing model is reformulated as convolution process of room impulse responses and sources \cite{cardoso1998blind, vincent2014blind}. The observed speech mixture is corrupted by reverberation and potential noise, which diminish the speech quality and increase the separation difficulty \cite{naylor2010speech}. Therefore, the performance of these close-talk separation methods is hard to fulfill the practical requirement. \emph{ii}) Most methods need to know the number of sources in the mixture and the output dimension is fixed to a specific source count. When the actual  number of sources mismatches the output dimension, termed as \emph{output dimension mismatch} problem \cite{huang2019research}, these methods are not facilitated to adjust their outputs. Furthermore, it's never easy to know the separated speech corresponds to which speaker in the mixture. Hence, in applications such as conference transcription, additional speaker verification or classification module is required to recognize the identity of separated speech \cite{rao2019target}, which we define as \emph{blind source allocation}.

First, to tackle with the first problem, far-field speech separation, it's a promising research direction to combine multi-channel speech separation (MSS) approaches with end-to-end approaches. MSS approaches include beamforming techniques based on microphone array signal processing \cite{gannot2017consolidated, adel2012beamforming,gannot2001signal,markovich2009multichannel}, blind source separation (BSS) \cite{sawada2006blind} and deep learning based methods \cite{chen2018multi, wang2019combining, wang2018integrating,wang2018multi,lianwu2019multi,yoshioka2018multi,drude2017tight}. With the direction of arrival (DOA) information, beamforming techniques conducts spatial filtering by means of the appropriate configuration of microphone array. The source from the desired direction is enhanced while interferences from other directions are suppressed. Since beamforming techniques employ spatial information to perform speech separation, when the sources are located closely to each other, these algorithms become less effective. BSS methods assume the sources are statistically independent to each other, and independent component analysis (ICA) is often used to model the separation process. However, when the reverberation is strong or the number of microphones is less than the number of sources, these methods degrade significantly. The deep learning-based methods employ the powerful feature learning capability of deep neural networks (DNN) to learn the T-F mask for the target source, using spatial binaural cues, e.g., interaural phase/time/level difference (IPD, ITD, ILD). The rationale lies in that, when sources are (W-) disjoint orthogonal, the binaural cues will form clusters within each frequency band for spatially separated directional sources with different time delays. This is also the theoretical basis of spatial clustering technique \cite{mandel2017multichannel, sawada2010underdetermined}. These spatial cues have been proven effective in deep learning-based frequency domain separation methods, especially when combined with spectral feature (e.g., logarithm power spectra, LPS) at input level \cite{wang2018multi,wang2018integrating,chen2018multi,lianwu2019multi,wang2019combining}.

Next, a common solution to output dimension mismatch and blind source allocation problem is to separate a target speaker once, instead of separating them all. To inform the network of which speaker needs to be extracted, target related prior knowledge needs to be incorporated. For instance, the voice characteristics of the target speaker. \cite{wang2018voicefilter,xu2018modeling,zmolikova2017speaker,vzmolikova2017learning,wang2018deep} employ speaker embeddings extracted from a reference signal from the target speaker to speaker-condition the separation system’s output. Apart from the target voice information, \cite{xiao2019single} proposed to utilize interference speakers’ embeddings as well, to increase the discrimination between speakers in the mixture. However, the main limitation for these speaker-aware methods is that they require the identity or the reference recordings from the target speaker. Other than the voice characteristics, direction information can be used to associate with a specific speaker in multi-channel speech separation. The direction information can either be estimated from acoustic/visual signals or predefined based on real usage scenario. Chen et al. \cite{chen2018multi} proposed a location-based angle feature which computes the cosine distance between the steering vector and inter-channel phase difference (IPD) for each source in mixture. But the phase ambiguities create difficulties for precisely discriminating one speaker from another in certain frequency bands. In \cite{wang2018spatial,wang2019combining}, Wang et al. developed two directional features to improve speech separation. One is the compensated IPD that shares the similar concept with angle feature and the other derives from beamforming outputs. Although impressive improvement is achieved with the directional features, \cite{wang2019combining} is a multi-stage system with high computational complexity. Also, it may neglect spatial ambiguity issue when speakers are close to each other, the directional features show less discrimination and it's hard to tell which speaker is corresponding to.

Taking two points discussed above into consideration, this study addresses the task of separating the target speech from multi-channel mixture in reverberant environments, assisted with directional information. We propose to fully exploit the spatio-temporal structure of multi-channel speech and integrate complementary separation cues into an end-to-end network for better performance while maintaining the real-time property.
First, we jointly model the temporal, spectral and spatial discriminability to create a more complete representation. The motivations are: 1) the separability of sources can be obtained from different views and they are complementary to each other. For instance, spectral sparsity in frequency domain, spatial diversity in spatial domain; 2) The discriminative capability of features from different domains depends on the conditions. For example, if sources come from close directions, then the spatial discrimination is not effective or even noisy. Under this condition, the model should rely on other features towards better separation.
Second, to reduce the latency, we adopt a purely end-to-end and single-pass separation network. It is a waveform-in, waveform-out separation system in a single neural network architecture, which inherited from Conv-TasNet. The model takes multi-channel mixture waveform and the directions of speaker(s) as input, and directly outputs estimated target speech waveform. We name this model as Temporal-Spatial Neural Filter, since it performs speech separation in the time-domain and separates the target speaker based on the informed directional information.

\begin{figure*}[t]
  \centering
  \includegraphics[width=\linewidth]{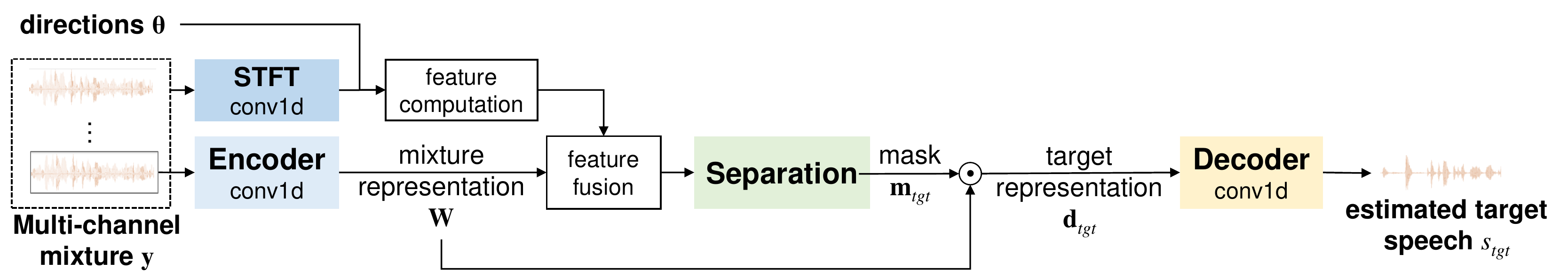}
  \caption{The framework of our proposed temporal-spatial neural filter. First, the multi-channel mixture waveform is passed to a conv1d layer that realizes the STFT function and produces multi-channel complex spectrogram. Meanwhile, the encoder transforms the reference channel's mixture $\mathbf{y}^{ref}$ to the mixture representation in a high-dimensional feature space. Second, along with the input directions $\mathbb{\phi}$, the feature computation module computes the spectral, spatial and directional features. Third, a joint acoustic representation is then obtained by concatenating the mixture representation and computed features in feature fusion module. Fourth, the separation module learns to estimate a mask for the target source, therefore the multiplicative product of the mask and mixture representation is the estimated target speech representation. Finally, the decoder reconstructs the target waveform from the masked mixture representation.
  }
  \label{fig:tsnf_framework}
\end{figure*}

The rest of paper is organized as follows. Section \ref{sec:physical_model} first introduces the physical model of far-field multi-speaker mixture. Then, section \ref{sec:single-channel_tss} reviews the monaural Conv-TasNet. Next, we elaborate on our proposed system in section  \ref{sec:proposed_system}, which is illustrated in figure \ref{fig:tsnf_framework}. We present the experimental setup and evaluation results in section \ref{sec:exp} and \ref{sec:result}, respectively, and conclude this paper in section \ref{sec:conclusion}.

\section{Physical Model}
\label{sec:physical_model}

In this study, we assume that the speakers do not move during speaking. The number of speakers and microphones are denoted as $C$ and $J$, respectively. The physical model for a reverberant mixture in time-domain is formulated as \cite{cardoso1998blind,vincent2014blind}:

\begin{equation}
    \mathbf{y}[n] = \underset{c=1}{\overset{C}{\sum}}{\mathbf{s}_c[n]}
    \label{eq:time_mixture}
\end{equation}
where $\mathbf{y}[n]=\left[ y^1[n],...,y^J[n] \right]^T$ is the $J$-dimensional vector of the mixture signal captured by the $J$-element microphone array, and $\mathbf{s}_c[n]=\left[ s_c^1[n],...,s_c^J[n] \right]^T=(\mathbf{h}_c\circledast x_c)[n]$ denotes the contribution of each source $c$ to microphone, where $\circledast$ is the convolution operator, $x_c$ is the dry clean speech of source $c$, $\mathbf{h}_c$ is the vector of room impulse responses associated with sound propagation from source $c$ to each microphone.
Formulate Eq. \ref{eq:time_mixture} in the time-frequency domain by means of the complex-valued STFT:
\begin{equation}
    \mathbf{Y}(t,f) = \underset{c=1}{\overset{C}{\sum}}{\mathbf{S}_{c}(t,f)}
    \label{eq:freq_mixture}
\end{equation}
where $\mathbf{Y}(t,f)$ and $\mathbf{S}_{c}(t,f)$ respectively denotes the complex spectrogram of mixture and reverberant image of source $c$ at time index $t$ and frequency band $f$.

In this study, given the multi-channel mixture signal $\mathbf{y}$, we aim to estimate individual signal $\hat{s}^{ref}_tgt$ for the target speaker, where $ref$ is the index of the reference microphone and $tgt$ is the target speaker's index in the mixture. It should be noted that our proposed system concentrate on speech separation task and do not take dereverberation into account. As a result, the model learns to estimate the reverberant speech $\hat{s}_c$ rather than dry clean speech $\hat{x}_c$.

\section{Time-domain Monaural Speech Separation}
\label{sec:single-channel_tss}

\begin{figure}[th]
    \centering
    \includegraphics[width=\linewidth]{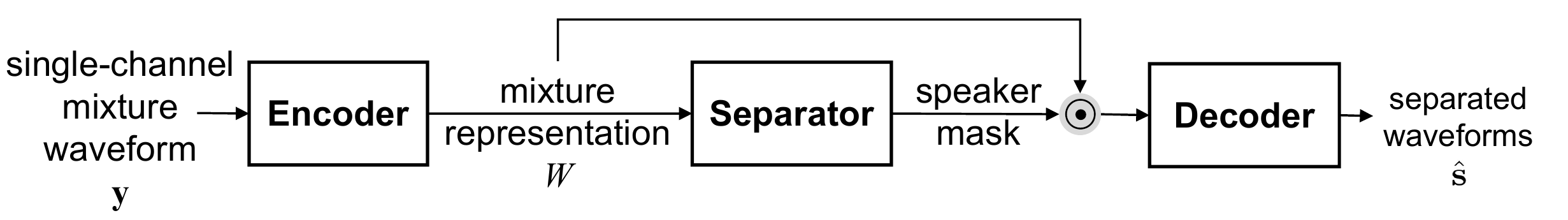}
    \caption{The diagram of Conv-TasNet.}
    \label{fig:tasnet}
\end{figure}

In this section, we will review Conv-TasNet \cite{luo2019convtasnet, luo2018surpass}, a deep learning framework for time-domain close-talk speech separation. Unlike the T-F masking based methods, Conv-TasNet replaces the STFT with a convolutional encoder-decoder architecture (figure \ref{fig:tasnet}).

Firstly, a short mixture segment $y$ is mapped to a high-dimensional representation $\mathbf{W}$ in the feature space by a linear encoder, which is a convolution 1d (conv1d) layer:
\begin{equation}
    \mathbf{W} = ReLU(y\circledast \mathbf{B})
    \label{eq:encoder}
\end{equation}
where $\mathbf{B} \in \mathbb{R}^{G\times L}$ is the encoder basis matrix, which contains $G$ convolution channels, each with window length $L$, $\circledast$ is the convolution operator, and $ReLU$ is rectified linear unit activation function. The number of convolution channels $G$ represents the number of basis functions. The kernel size $L$ and stride are the window length and hop size, respectively.
Then, the separation module adopts a temporal fully-convolutional network (TCN) \cite{lea2016temporal, feichtenhofer2016convolutional}, which computes a mask $\mathbf{M}_c$ for each source $c$, similar to the T-F masking. As a result, the speaker representation can obtain by a multiplicative product $\mathbf{D}_c=\mathbf{M}_c \odot \mathbf{W}$. The TCN consists of stacked dilated conv1d blocks, where each layer in the block features with exponentially increasing dilation factors. Meanwhile, in these dilated conv1d blocks, the traditional convolution is substituted with depthwise separable convolution to further reduce the parameters.

Finally, the decoder invert the speaker's representation $\mathbf{D}_c$ back to the time-domain signal, using 1d linear deconvolution:
\begin{equation}
    \hat{x}_c = \mathbf{D}_c\mathbf{V}
    \label{eq:decoder}
\end{equation}
where $\mathbf{V} \in \mathbb{R}^{G\times L}$ is the decoder basis matrix. To optimize the network, instead of using a time-domain mean squared error (MSE) loss, the speech separation metric scale-invariant signal-to-distortion (SI-SDR) is used to directly optimize the separation performance, which is defined as:

\begin{equation}
\left\{
\begin{array}{lr}
x_{\text{target}}:=\frac
{\left<\hat{x}, x\right>x}
{\left\|x\right\|_{2}^{2}} \\
e_{\text{noise}}:=\hat{x}-x_{\text{target}}  \\
\text{SI-SDR}:=10\log_{10}\frac
{\left\|x_{\text{target}}\right\|_{2}^{2}}
{\left\|e_{\text{noise}}\right\|_{2}^{2}}
\end{array}
\right.
\label{eq:si_sdr}
\end{equation}
where $x$ and $\hat{x}$ are the dry clean and estimated source waveform, respectively. The zero-mean normalization is applied to $x$ and $\hat{x}$ to guarantee the scale invariance.

\section{Proposed system}
\label{sec:proposed_system}

The proposed model is an integrated waveform-in waveform-out separation system in a single neural network architecture. The model consists of six processing modules: STFT conv1d layer, encoder, feature computation, feature fusion, separation and decoder. We have elaborated the encoder, separation module and decoder in section \ref{sec:single-channel_tss}, all of which inherited from Conv-TasNet. In this section, we will give the details of rest modules. First, to enable end-to-end training, the STFT operation is implemented with convolution layer and used to extract frequency domain features, described in Sec \ref{subsec:STFT_layer}; Second, the formulation and online computation of spatial and directional features are respectively presented in \ref{subsec:SI_features} and \ref{subsec:SR_features}. Finally, the fusion details are elucidated in \ref{subsec:fusion}.



\subsection{STFT convolution layer}
\label{subsec:STFT_layer}

Given a window function $w$ with length $N$, the spectrum $Y$ can be calculated by standard STFT:
\begin{equation}
\begin{split}
y[n] \xrightarrow[]{\tt STFT} Y(t,f)
&=\overset{N-1}{\underset{n=0}{\sum}}y[n]w[n-t]\exp{\left(-i\frac{2 \pi n}{N}f\right)}
\end{split}
    \label{eq:STFT}
\end{equation}
To speed up the separation procedure, we intend to implement all the feature computation with network layers and operations. Following \cite{wichern2018phase, gu2019end}, we reformulate the standard STFT as a convolution kernel in Eq.~\ref{eq:STFT}. To distinguished with the standard STFT, we use $\tau$ and $m$ as the frame and frequency band index, which is actually the feature map height and width index after convolution, respectively:

\begin{equation}
y[n] \xrightarrow[]{\tt STFT} Y_{\tau,m}
=\overbrace{e^{-i\frac{2\pi \tau}{N}m}}^{\tt{phase~factor}} (y[\tau] \circledast \overbrace{w[\tau]e^{-i\frac{2\pi \tau}{N}m}}^{\tt{STFT~kernel}} )
    \label{eq:STFT_conv}
\end{equation}
Note the phase factor in Eq.~\ref{eq:STFT_conv} is constant vector, which means it will neither affect the magnitude ($|\cdot|=1$) nor IPD. The complex STFT kernel can be splited in real and imaginary parts:

\begin{equation}
\begin{split}
\mathbf{K}^{\tt real}_{\tau,m} &=w[\tau] \cos(2\pi \tau m/N) \\
\mathbf{K}^{\tt imag}_{\tau,m} &=-w[\tau] \sin(2\pi \tau m/N)
\end{split}
    \label{eq:STFTkernel}
\end{equation}

The shape of the kernel is determined by the $w[n]$. The size of the kernel is actually the window length $N$ of $w[n]$. The stride of convolution equivalents to the hop size in the STFT operation. We can also customize the number of kernels and use preferred kernel sizes other than $N$. The stride in convolution is also now configurable other than just using the hop size of STFT. In this work, to match the encoder configuration, the length and stride of STFT kernel is set the same as encoder kernel length $L$ and stride.
The reference channel's LPS is served as the spectral feature, calculated by:
\begin{equation}
\textit{LPS}_{\tau,m}=10\log \left (
\left ( \mathbf{y}^{ref} \circledast \mathbf{K}^{\tt real}_{\tau,m} \right )^2 +
\left (\mathbf{y}^{ref} \circledast \mathbf{K}^{\tt imag}_{\tau,m} \right )^2
\right )
    \label{eq:lps}
\end{equation}

\subsection{Enhancing Speech Separation with spatial features}
\label{subsec:SI_features}

As discussed in introduction, well-established spatial cues like IPDs have shown great beneficial for T-F masking based MSS methods \cite{lianwu2019multi,wang2018multi,chen2018efficient,wang2018spatial}.  The standard IPD is computed by the phase difference between channels of complex spectrogram as:
\begin{equation}
\textit{IPD}^{(u)}(t,f)=\angle\mathbf{Y}^{u_1}(t,f)-\angle\mathbf{Y}^{u_2}(t,f)
\label{eq:ipd_ori}
\end{equation}where $u_1$ and $u_2$ represents two microphones' indexes of the $u$-th microphone pair.
In our study, given the STFT kernel $\mathbf{K}$, the $u$-th pair of IPD can be computed by:
\begin{equation}
\label{eq:kernel_IPD}
\textit{IPD}^{(u)}_{\tau,m}=\arctan\left( \frac{\mathbf{y}^{u_1} \circledast \mathbf{K}^{\tt real}_{\tau,m}  }{\mathbf{y}^{u_1} \circledast \mathbf{K}^{\tt imag}_{\tau,m}  }\right)
-\arctan\left(
\frac{\mathbf{y}^{u_2} \circledast \mathbf{K}^{\tt real}_{\tau,m}  }{\mathbf{y}^{u_2} \circledast \mathbf{K}^{\tt imag}_{\tau,m}  }
\right)
\end{equation}

\subsection{Target speech separation with directional features}
\label{subsec:SR_features}

\begin{figure*}[th]
  \centering
  \includegraphics[width=14cm]{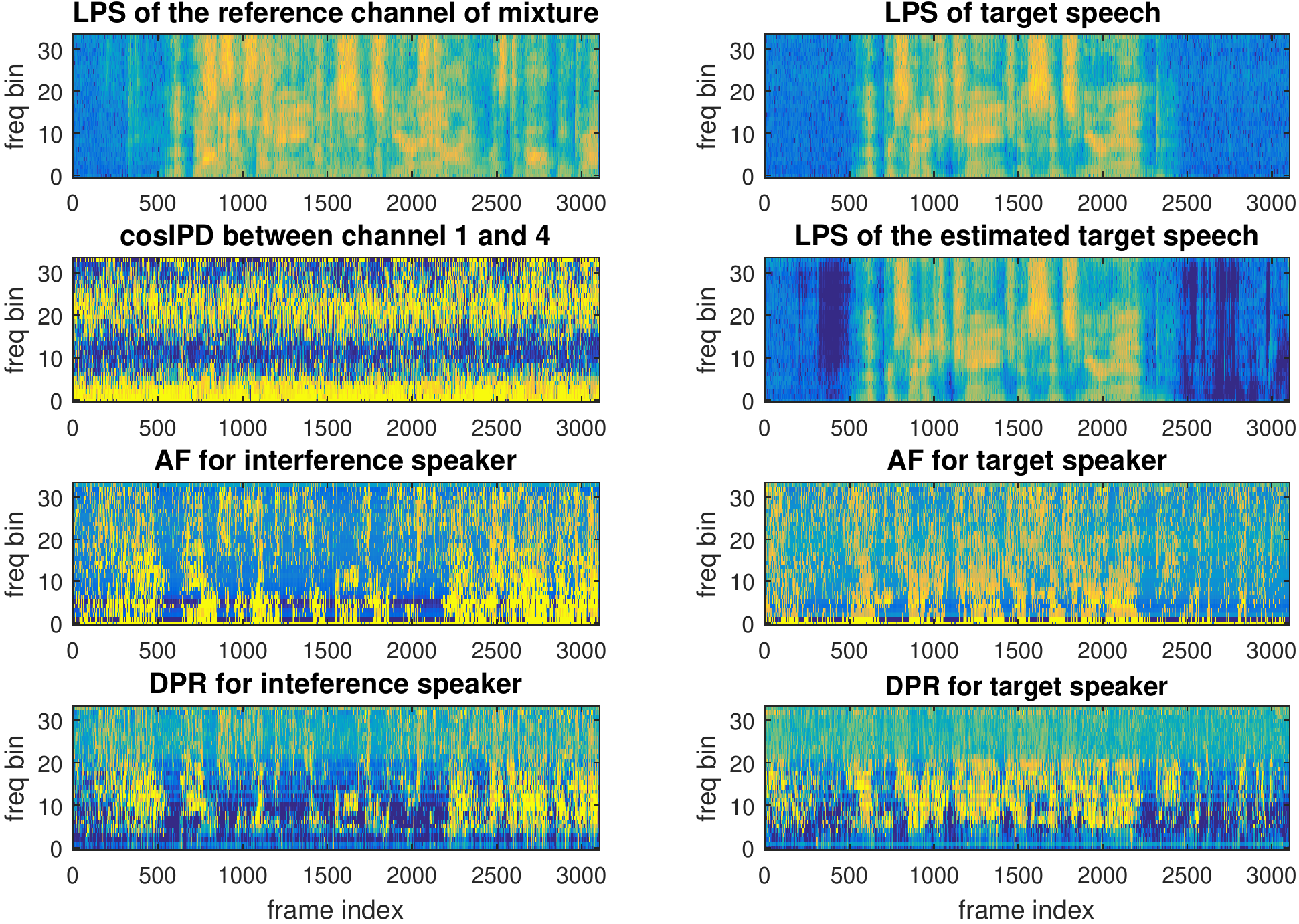}
  \caption{A example in WSJ0 2-mix of the mixture logarithm power spectrum (LPS), directional features DPR and AF for the target speaker's direction and the interference speaker's direction, the LPS of ground truth target speech and the LPS of the recovered target speech. }
  \label{fig:all_features}
\end{figure*}

Spatial feature such as IPD successfully extracts the spatial information of all the sources in the mixture signal. Moreover, with some or all of the speaker directions, features of specific speaker-dependent direction could be extracted to improve the performance of separation further. In this paper, it is assumed that the oracle location of each speaker is known by the separation system. This is an reasonable assumption in some real applications, for example, the speaker location could be detected by face detection techniques with very high accuracy.
In this work, we use an improved version of angle feature (AF) \cite{chen2018multi} and our proposed directional power ratio (DPR) as directional features.

The location-guided angle feature (AF) is first introduced in \cite{chen2018multi}, specially designed for the seven-element circular microphone array. We reformulate AF so that it can be applied to general microphone array topology. AF measures the cosine distance between the steering vector, which is formed according to the direction of the target speaker, and IPDs:
\begin{equation}
\textit{AF}_{\phi,\tau,m}=Real \left (
\overset{U}{\underset{u=1}{\sum}}
\frac{
\mathbf{a}_{\phi,m}^{u}
\exp(\textit{IPD}^{u}_{\tau,m})
}
{
\left |\mathbf{a}_{\phi,m}^{u}
\exp(\textit{IPD}^{u}_{\tau,m}) \right |
}
\right )
\label{eq:AF}
\end{equation}
where $\mathbf{a}_{\theta,m}^{u1}$ is the steering vector coefficient for target speaker from $\phi$ at the $m$-th frequency band for first microphone of $u$-th pair. Also, the pre-masking step in \cite{chen2018multi} is also applied to increase the discrimination of AF. The design principle of AF lies in that if the T-F bin is dominated by the source from desired direction, then the similarity between the steering vector and IPD will be close to 1, otherwise close to 0.
This feature provides the desired speaker's directional information to the network so that the network is expected to attend to the target speech. However, since the steering vector is computed with the exact $\phi$, the tiny estimation error may lower the quality of AF and cause the degradation of separation performance.

In our previous study, DPR was defined based on the output power of multi-look Cardioid Filters that carefully designed for a specific microphone array. However, the design can be complex and not easy to implement inside the network. In this work, for simplicity and reproducibility, we replace the Cardioid filters with a set of delay-and-sum beamformers (DAS-BF) steered at different directions. For a given microphone array and a pre-defined direction grid $\{\theta_1,\theta_2,...\theta_P\}$, a set of DAS-BFs are denoted as $\mathbf{w}_{p,m}\in\mathbb{C}^{J}$, which aims to enhance sound sources from direction $\theta_p$ for $m$-th frequency band.
Under far-field and free-field conditions, the DAS-BF steered at $\theta_p$ is computed by

\begin{equation}
    \label{eq:ds_bf}
    \mathbf{w}_{p,m} = \exp(-2\pi im \triangle t_{\theta_p,j}) / J
\end{equation}
where
$\triangle t_{\theta_p,j}$
denotes the time difference of arrival (TDOA) from microphone $ref$ to $j$ when the source is from direction $\theta_p$.
Assuming these beamformers can provide well enough spatial separation and the multiple speakers are not closely located in the space, we can use the processing output power of $\mathbf{w}_{p,m}$ as a reasonable estimation of the signal power from direction $\theta_p$. Therefore, the DPR can be considered as an indicator of how well is a T-F bin $(\tau,m)$ dominated by the signal from direction $\theta_p$, defined as follows:
\begin{equation}
 \textit{DPR}_{\theta_p,\tau, m}=\frac{
\left \|\mathbf{w}_{p,m}^{H}\mathbf{Y}_{\tau, m}  \right \|_{F}^{2}
}
{
\sum_{p'=1}^{P}
{
\left \|\mathbf{w}_{p',m}^{H}\mathbf{Y}_{\tau, m}  \right \|_{F}^{2}
}
}
  \label{eq:dpr}
\end{equation}
where $\mathbf{Y}_{\tau, m}$ is the complex spectral vector of multi-channel mixture signal in T-F bin $(\tau,m)$, computed as $\mathbf{Y}_{\tau, m} = \mathbf{y} \circledast \mathbf{K}^{\tt real}_{\tau,m} + i * ( \mathbf{y} \circledast \mathbf{K}^{\tt imag}_{\tau,m} )$. In our implementation, the beamformer coefficients of $\mathbf{w}_p$ are set as weights of a linear layer. As a result, the DPR feature can be calculated by a set of linear layers and safe divide operation.

Although the resolution of DPR, i.e., $\theta_{p+1}-\theta_{p}$, is lower than that of AF, one significant advantage is that it is more robust to the direction estimation error since it covers a particular range of directions. Therefore, combining AF and DPR as directional features can be a good choice to promise both the robustness and precision for separation.

Figure \ref{fig:all_features} illustrates the directional features when applied to a sample in spatialized dataset WSJ0 2-mix. It can be interpreted from the figure that both reformulated AF and the proposed DPR that focus at target direction can provide clues for the target source's contribution. Although DPR's saliency is weaker on lower frequency bands, the model can refer to AF as supplementary clues. Also, discrimination between target and interference speech can be achieved by combining the directional features for target and interference speaker.

\subsection{Feature fusion}
\label{subsec:fusion}

Before fed into the separation module, the mixture representation $\mathbf{W}$ and computed spectral, spatial, directional features are fused to form a joint acoustic representation. Specifically, with the same convolution kernel size and stride, the numbers of time steps of $\mathbf{W}$, $\textit{LPS}$, $\textit{IPD}$, $\textit{AF}$ and $\textit{DPR}$ are the same. Then, these features are concatenated along the feature axis and passed to the proceeding layers.

It should be noted that, in this work, we have tried to introduce an attention module to automatically learn the time-varying contribution of each feature, like our previous trial \cite{gu2019neural}. However, the gain is minor and extra network parameters are introduced. The reason may lie in that the first layer in the separation module is a conv1$\times$1 (fully-connected) layer that has associated the weight with each feature. Also, the learned contribution of directional features is always high. Therefore, we do not report the corresponding results in this paper and leave it to discussion.


\section{Experiments procedures}
\label{sec:exp}

\subsection{Dataset}
\begin{figure}[h]
    \centering
    \includegraphics[width=3cm]{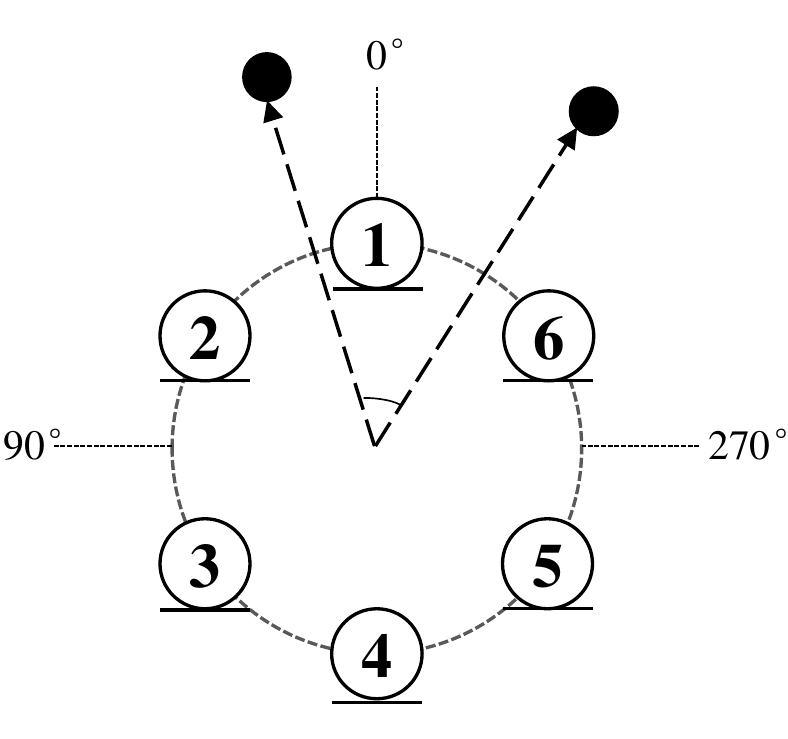}
    \caption{The microphone configuration and space grid for the experiments.}
    \label{fig:mic}
\end{figure}

We simulated a spatialized reverberant dataset derived from Wall Street Journal 0 (WSJ0) 2-mix and 3-mix corpus, which are open and well-studied datasets used in monaural and multi-channel speech separation \cite{hershey2016deep,yu2017permutation,luo2019convtasnet,lianwu2019multi,wang2019combining}. In 2-mix corpus, there are 20,000, 5,000 and 3,000 multi-channel, reverberant, two-speaker mixed speech in training, development and test set respectively. The amount of three-speaker speech mixture in training, development and test set of 3-mix corpus is respectively the same as that in 2-mix corpus. The performance evaluation is all done on test set, the speakers in which are all unseen during training. The mixing signal-to-noise ratio (SNR), pairs, dataset partition are exactly coincident with anechoic monaural WSJ0 2-mix and 3-mix. For the selection of microphone array, we consider a 6-microphone circular array of 7cm diameter with speakers and the microphone array randomly located in the room, as illustrated in figure \ref{fig:mic}. The two speakers and microphone array are on the same plane and all of them are at least 0.3m away from the wall. The image method [23] is employed to simulate RIRs randomly from 3000 different room configurations with the size (length-width-height) ranging from 3m-3m-2.5m to 8m-10m-6m. The reverberation time T60 is sampled in a range of 0.05s to 0.5s.
For 2-mix spatialized dataset, samples with angle difference between two simultaneous speakers of 0-15$\degree$, 15-45$\degree$, 45-90$\degree$ and 90-180$\degree$ respectively account for 16\%, 29\%, 26\% and 29\%. The angle difference between two speakers is defined as: $ad(\phi_1, \phi_2)=\min(|\phi_1-\phi_2|, |360\degree-\phi_1+\phi_2|)$, where $\phi_1$ and $\phi_2$ respectively denotes the direction (in degree) of two speakers. For 3-mix spatialized dataset, the angle difference is defined as the smallest angle difference between the target speaker and other interference speakers. The proportion for each angle difference range is 29\%, 36\%, 23\% and 12\%. When there is more than one interference speaker, the angle difference is defined as the degree difference between the target speaker and the closest speaker, i.e., $ad({\phi_1, ..., \phi_C})=\min_{c,c\ne tgt}(ad(\phi_{tgt}, \phi_{c}))$.

\subsection{Feature extraction and Network hyper-parameters}

The selected pairs for IPDs are (1, 4), (2, 5), (3, 6), (1, 2), (3, 4) and (5, 6) in all experiments. These pairs are selected considered that the distance between each pair is either the furthest or nearest.

The reference channel is set to the first channel of speech mixture waveform as default. To match the encoder, all frequency domain features, including LPS, IPDs and directional features, are extracted with 2.5ms window length and 1.25ms hop size with 64 FFT points. That means that the length and stride of STFT kernel is the same as encoder filter length $L$ and stride, i.e., 40 and 20 points. The number of filters is set to 33 since we round the window length to the closest exponential of 2, i.e., 64.

For DPR computation, we use 36 fixed spatial filters and the $p$-th filter is steered at azimuth $10p\degree$. This resolution is selected empirically considering the balance between precision and robustness.

All the data is sampling at 16kHz. All hyper-parameters are the same with the best setup of Conv-TasNet in \cite{luo2018surpass}, except $L$ is set to 40 and encoder stride is 20. Batch normalization (BN) is adopted because it has the most stable performance. Note that we do not adopt the update version of Conv-TasNet \cite{luo2019convtasnet}, although it exhibits better performance and smaller model size. Also, adopting global layer normalization (gLN) instead of BN could further improve the performance a lot. Since our target is to approach real-time separation, we prefer the early version of Conv-TasNet without densely connected structure and global normalization.

\subsection{Training paradigms}

For the output of separation system, there are two cases: i) the system outputs the estimated speech of all speakers in the mixture. When there is no directional information, permutation invariant training \cite{yu2017permutation} is adopted to tackle with label permutation problem. This means at the inference, the output-speaker assignment is unknown and additional speaker recognition procedure is required. While all speakers' directions are given, the output speech order is exactly the order of input speakers' directions. Therefore, the target speech can be obtained by selecting the corresponding index. However, under small angle angle range, i.e, $<15\degree$, the output order become vague due to spatial ambiguity issue. For 2-targets case, we evaluate the performance with the permutation that achieves the best result.
ii) Only the target speech is estimated. For this 1-target case, we investigate two types of input: 1) only target speaker's directional feature is provided as target-related information, denoted as \emph{tgt}. This system is more applicable in practical since it can perform target speech separation with only the target speaker's direction; 2) Following our recent study \cite{gu2019neural} and \cite{xiao2019single}, interference speaker's directional feature is also attached to the target speaker's to enhance the discrimination between different directions, denoted as \emph{tgt+intf}. For target speech separation task, the performance is obtained by running the separation network by the times of speaker number, each time selecting one speaker in mixture signal as target.

SI-SDR (Eq. \ref{eq:si_sdr}) is utilized as training objective. The training uses chunks with 4.0 seconds duration. The batch size is set to 32.

\subsection{Evaluation metrics}

Following the recent advances in speech separation metrics \cite{le2019sdr}, average SI-SDR is adopted as the main evaluation metric. Also, following the common practice, average SDR computed using BSS\_EVAL toolbox \cite{vincent2006performance} and perceptual estimation of speech quality (PESQ) are used to measure the speech quality and intelligibility. Since we do not perform speech dereverberation, we consider the reverberant image of source $s^{ref}_c$ for metric computation.

Instead of the overall performance, we also report the performances under different ranges of angle difference between speakers. A good model should perform excellent overall and balanced under all angle difference ranges. We intend to investigate the relative performance difference under different ranges and give a more comprehensive assessment for the model.

\section{Results and analysis}
\label{sec:result}

\subsection{Window length selection for frequency domain features}
As discussed in \ref{subsec:STFT_layer}, the features that extracted in frequency domain using different analysis window length and hop size may mismatch the encoder output.
Considering the short-term stationary characteristics of speech, the typical window length for T-F masking based methods is 15-32ms (240-512 FFT points for speech sampled at 16kHz). Also, a larger window length can provide higher frequency resolution for more refined spectrum analysis. However, using this window length, the time resolution is much lower than that of mixture representation output by the encoder, i.e., 2.5ms (40 points). Therefore, we investigate the window length for extracting the frequency domain features that works the best with encoded representation. To ensure the synchronization between frequency domain features and the mixture representation, upsampling operation is applied to the frequency domain features when the stride (hop size) is larger than 40.

\begin{table}[b]
  \caption{SDRi (dB) and SI-SDRi (dB) performances of 2-target speech separation with different FFT size, window length and stride (hop size) on far-field WSJ0 2-mix.
}
  \label{tab:fft_size}
  \centering
  \begin{tabular}{c|ccc|cc}
    \hline
    \textbf{Features} & \textbf{FFT} & \textbf{Win.} & \textbf{stride} & \textbf{SI-SDRi (dB)} & \textbf{SDRi (dB)} \\
    \hline
    Conv-TasNet  & - & 40 & 20 & 9.1  & 9.4   \\
    \hline
    +cosIPD & 512 & 512 & 256 & 10.1  & 10.5 \\
    +cosIPD  & 256 & 256 & 128 & 10.6 & 11.0\\
    +cosIPD  & 128 & 128 & 64 & 10.8 & 11.2\\
    +cosIPD  & 64 & 40 & 20 & \textbf{11.2} & \textbf{11.6}\\
    +cosIPD  & 40 & 40 & 20 & 9.1 & 9.5\\
    \hline
  \end{tabular}
\end{table}

Table \ref{tab:fft_size} reports the results. As we can observe, the configuration of 64-point FFT along with window length of 40 achieves the best performance, since its window length and stride is coincident with those of encoder. In the following experiments, we all adopt the configuration of 64-point FFT, 2.5ms window length and 1.25ms hop size for computing frequency domain features.

\subsection{Ablation study on feature combination}

\begin{table*}[t]
  \caption{SDRi (dB) and SI-SDRi (dB) performances of target separation systems on far-field WSJ0 2-mix.}
  \label{tab:ablation}
  \centering
  \begin{tabular}{c|l|cccc|c|c}
    \hline
    \multirow{2}{*}{\textbf{\# of target}} &
    \multirow{2}{*}{\textbf{Features \& Setup}} &
    \multicolumn{5}{c}{\textbf{SI-SDRi (dB)}} &
    \multirow{2}{*}{\textbf{SDRi (dB)}} \\
    & &$<$15\degree &15\degree-45\degree &45\degree-90\degree &$>$90\degree & Ave. \\
    \hline
    2  & Single-channel Conv-TasNet  & 8.5 & 9.0  & 9.1	& 9.3	&9.1   & 9.4     \\
    \hline
    2   & +LPS & 9.2 & 9.7 & 9.6 & 10.0 & 9.7 & 10.0 \\
    2    & +cosIPD  & 8.5 & 11.8 & 12.0	& 11.6	&11.2   & 11.6   \\
    2   & +cosIPD,sinIPD  & 7.7 & 11.6 & 12.3 & 12.6 & 11.5 & 11.9\\
    2  & +LPS, cosIPD, sinIPD &8.1 &11.9 &12.7 &13.1 & 11.9 & 12.2\\
    \hline
    2   & +cosIPD, AF  &10.7	&12.9	&13.4	&13.7	&12.9 & 13.3\\
    2 & +cosIPD, DPR & 9.2 & 12.8 & 13.3 & 13.6 & 12.6 & 13.0\\
    2 & +cosIPD, AF, DPR & 10.5 & 13.0 & 13.5 & 13.8 & 13.0 & 13.4 \\
    2 & +cosIPD, sinIPD, AF, DPR & 10.6 & 13.1 & 13.6 & 13.9 &\textbf{13.1} & \textbf{13.4}\\
    \hline
    1 & +cosIPD, AF (tgt) & 9.6 & 12.9 & 13.3 & 13.6 & 12.7 & 13.1\\
    1& +cosIPD, AF (tgt+intf) & 9.8 & 12.9 & 13.4 & 13.7 & 12.8 & 13.2\\
    1& +cosIPD, DPR (tgt+intf) & 5.3 & 12.5 & 13.0 & 13.3 & 11.8 & 12.2\\
    1&+cosIPD, AF, DPR (tgt) & 9.4 & 12.7 & 13.1 & 13.5 & 12.5 & 12.9\\
    1&+cosIPD, AF, DPR (tgt+intf) & 9.8 & 13.0 & 13.5 & 13.8 &\textbf{12.9} & \textbf{13.3} \\
    1& +cosIPD, sinIPD, AF (tgt+intf) & 9.5 & 12.5 & 13.0 & 13.3 & 12.4 & 12.8\\
    \hline
  \end{tabular}
\end{table*}

In order to evaluate the contribution of different features and their combinations, we conduct an ablation study on the feature selection on the task of speech separation from a mixture of two  speakers. 

Table \ref{tab:ablation} shows the results of our ablation study. The table includes evaluation using SI-SDRi and SDRi with three input setups: 1) temporal encoded representation only (Single-channel Conv-TasNet); 2) adding spectral (LPS) and spatial features (IPD); 3) adding spectral, spatial and directional features (DPR, AF).

The single-channel 2-speaker separation network performs with SI-SDRi of 9.1dB. Adding the LPS of the mixture's reference channel elevates the performances under all angle difference ranges for about 0.7dB. Further adding spatial features (cosIPD and sinIPD) boosts the overall performance to 11.9dB. Particularly, the inclusion of spatial features greatly enhances the performance under large angle difference range for about 2.8dB.
With the aid of directional features, better performance can be achieved compared to systems only with spatial and spectral features, i.e., 13.1dB versus 11.9dB.

As for single target separation, compared to feeding both target's and interference's AF, the performance of model that only provides target's AF drops only 0.1dB. Adding DPR feature further boost overall performance, especially under large angle difference range. However, the inclusion of sinIPD may weaken the contribution of AF and reduces the performance.

\subsection{Varying mixing number}

To illustrate the ﬂexibility of our proposed model, in table \ref{tab:any_mix}, we summarize the single target separation performance of the models respectively trained on WSJ0-2mix, WSJ0-3mix and both WSJ0-2mix and WSJ0-3mix. The computed features are cosIPD, AF and DPR (tgt+intf). All the models are tested on both two- and three-speaker mixtures without knowing the source number in the mixture.

For 3-speaker mixture, the interference speakers’ directional features are merged as one, either selecting the closer interference speaker to target speaker, or summing/averaging the directional features of both interference speakers. In this experiment, the closest interference speaker's direction is used to compute the interference speaker's directional feature.
For the model trained on only 2-speaker mixtures to perform 3-speaker separation, we always chose the closest interference speaker to the target speaker to compute the directional feature.

Compared with models that trained with the dataset mixed by a fixed number of speakers, the model trained with both 2- and 3-mix dataset shows superior performance both under 2-speaker and 3-speaker mixing condition. This result indicates that a single model can handle a varying and unknown number of speakers' mixture. This is of great practical importance, since a priori knowledge about the number of speakers is not needed at test time.

\begin{table}[t]
  \caption{SDRi (dB) and SI-SDRi (dB) performances of single target speech separation on far-field WSJ0 2-mix and 3-mix. The extracted features are cosIPD, AF, DPR (\emph{tgt+intf}) for all models.
}
  \label{tab:any_mix}
  \centering
  \begin{tabular}{c|cc|cc|cc}
    \hline
    \multirow{2}{*}{\textbf{Tr. Dataset}} &
    \multicolumn{2}{c}{\textbf{SI-SDRi (dB)}} &
    \multicolumn{2}{c}{\textbf{SDRi (dB)}} &
    \multicolumn{2}{c}{\textbf{PESQ}} \\
    & 2spk &3spk &2spk &3spk &2spk &3spk\\
    \hline
    Mixture & 0 & 0 & 0 & 0 & 1.98 & 1.62\\
    \hline
    2-mix    & 12.9 & 5.9 & 13.3 & 6.3 & - & -\\
    3-mix  & 11.7 & 9.5 & 12.0 & 9.8 & - & -\\
    2+3-mix  & \textbf{13.5} & \textbf{11.1} & \textbf{13.9} & \textbf{11.5} &3.28 &2.74\\
    \hline
    IBM & 12.2 & 12.7 & 12.6 & 13.1 & 3.11 & 2.70\\
    IRM & 12.0 & 12.6 & 12.4 & 13.0 & 3.79 & 3.66\\
    IPSM & 15.0 & 15.8 & 15.4 & 16.3 & 3.95 & 3.84\\
    \hline
  \end{tabular}
\end{table}

\subsection{Comparisons with other approaches}

We take two types of rival systems into consideration:
1) monaural end-to-end system with state-of-the-art performance, i.e., single-channel Conv-TasNet \cite{luo2018surpass};
2) deep learning-based MSS methods, including Freq-BLSTM based speech separation methods, multi-channel deep clustering (DC) \cite{wang2018multi} and neural spatial filter \cite{gu2019neural}.
For all the Freq-BLSTM based methods, the input is the concatenation of cosIPD and LPS, and the output is the T-F mask for each source. When the training objective is phase-sensitive spectrum approximation (PSA), the estimated mask is referred as phase-sensitive mask (PSM), which is a commonly used training target in speech separation task \cite{kolbaek2017multitalker, wang2014training}. We also investigate the end-to-end training objective, SI-SDR, which is the same as that of our model. MISI-5 proposed by \cite{wang2018end} is a phase reconstruction algorithm that implemented with iterative network layers. As for multi-channel DC, it integrates conventional spatial clustering with DC \cite{hershey2016deep} by including IPD patterns in the input of deep clustering network. First, the 2-channel DC model is trained on interaural patterns extracted from different microphone pairs. Then, the embeddings of each pair that produced by the DC network is stacked along the embedding dimension. Finally, perform K-means on these embeddings and obtain the estimated T-F binary mask. Neural Spatial Filter takes LPS, cosIPD, DPR, Directional signal-to-noise ratio (DSNR) and AF as the LSTM network input, where both DPR and DSNR are calculated by Cardioid Filters. The network is trained with spectrum approximation objective to estimate a ideal ratio mask (IRM).

Table \ref{tab:all} also report the performances of oracle masks for reference, including ideal binary mask (IBM), IRM and ideal PSM (IPSM) \cite{wang2014training}. These masks are computed with 256-point FFT, 16ms hanning window. The processing speed is evaluated as the average processing time for the systems to separate each frame in the mixtures, which referred as time per frame (TPF) \cite{luo2018surpass}. If a system can be implemented in real time, the required TPF should be smaller than the frame length. For the CPU configuration, we tested all the systems with one processor on an Intel Xeon E5-2680 CPU. For the GPU configuration, the systems are tested on one Tesla M40 GPU.
Specially, the frame length for both single-channel Conv-TasNet and our proposed temporal-spatial neural filter is 2.5ms. While for other deep learning based multi-channel separation methods, the frame length is 32ms.

Note that we do not include a state-of-the-art multi-channel speech separation system \cite{wang2019combining} for comparison, which also proposes to enhance the separation with directional features. The main reason is that it is a multi-stage system that adopts the strategy of \emph{Separate-Localize-Enhance}. The products of separation stage are also utilized in enhancement stage. Also, authors claimed that this system focused on offline processing so that the computational cost may be high.

As can be observed in Table \ref{tab:all}, our proposed Temporal-Spatial Neural Filter exhibits the best performance among all the comparison methods under all the angle difference ranges. Also, our model's size is smaller and the processing speed is faster, which means it is a more promising approach for real-time applications. With the same end-to-end training objective and similar network structure, our method outperforms single-channel Conv-TasNet by 4.4dB of SI-SDR. This demonstrates the benefits of the joint representation of temporal, spectral, spatial and directional features. With the similar input features, the performance of end-to-end Temporal-Spatial Neural Filter obtains 4.4dB improvement over the LSTM-based Neural Spatial Filter. Furthermore, the processing speed is greatly reduced, which indicates the effectiveness of on-the-fly feature extraction and end-to-end network.

\begin{table*}[h]
  \caption{Performances of target separation systems on simulated far-field WSJ0 2-mix test set.}
  \label{tab:all}
  \centering
  \begin{tabular}{c|c|l|c|cc|c|c}
    \hline
    \multirow{2}{*}{\textbf{\# params}} &
    \multirow{2}{*}{\textbf{\# mic.}}&
    \multirow{2}{*}{\textbf{Approach \& Setup}} &
    \multirow{1}{*}{\textbf{CPU/GPU}} &
    \multicolumn{3}{c}{\textbf{SI-SDRi (dB)}} &
    \multirow{2}{*}{\textbf{SDRi (dB)}}\\
    & & &\textbf{TPF \#ms} &$<$15\degree &$>$15\degree & Ave. &  \\
    \hline
    -  & - & Mixture  &  - & 0 & 0 & 0 & 0  \\
    \hline
    8.8M & 1 & Single-channel Conv-TasNet \cite{luo2019convtasnet} & 0.4 / 0.02 & 8.5 & 9.1  & 9.1	& 9.5   \\
    21.7M & 6 & Multi-channel Freq-BLSTM (PSA)  & 22 / 8.0 & 6.5 & 9.8 & 9.3 & 9.7   \\
    30.0M & 6 & Multi-channel Freq-BLSTM (SI-SDR) & 22 / 8.0 & 6.5 & 10.1 & 9.5 & 9.9   \\
    30.2M & 6 & Multi-channel Freq-BLSTM (SI-SDR+MISI-5) & 24 / 8.2 & 7.3 & 10.5 & 10.0 & 10.5   \\
    33.8M & 6 & Multi-channel DC* \cite{wang2018multi} &39 / 14 &9.1 & 10.1 & 9.9 & 10.3  \\
    \hline
    21.5M & 6 & Neural Spatial Filter \cite{gu2019neural} &23 / 7.7 &4.8 & 10.1 & 9.1 &9.5   \\
    8.8M & 6 & Temporal-Spatial Neural Filter (proposed) &0.5 / 0.03 &\textbf{10.8} &\textbf{14.0} &\textbf{13.5} &\textbf{13.9} \\
    \hline
    - & -  & IBM  & - & 12.1 & 12.1 & 12.2  & 12.6	   \\
    - & -  & IRM  &- & 12.0 &12.0  &12.0  & 12.4     \\
    - & -  & IPSM  &- & 15.0 &15.0  &15.0  & 15.4     \\
    \hline
  \end{tabular}
\end{table*}

\subsection{Sensitivity of proposed system to direction estimation error}

In above experiments, we assume the oracle direction of each speaker is known without any estimation error. However, in real-world scenarios, the DOA estimation algorithms or face location techniques may suffer from interferences from acoustic and visual signals and bring about the estimation error. Therefore, in order to investigate how the error affects the proposed model's performance, we introduce random error to the input target speaker's direction. Specifically, we deviate the ground truth target direction for ±1 to ±10$\degree$ for all the testing samples of WSJ0 2-mix and 3-mix, where the positive or negative deviation is random. We compare two systems that integrated with different directional features, one with AF only, the other one with AF and DPR. All the directional features are computed using deviated directions.

\begin{figure}[h]
    \centering
    \includegraphics[width=7cm]{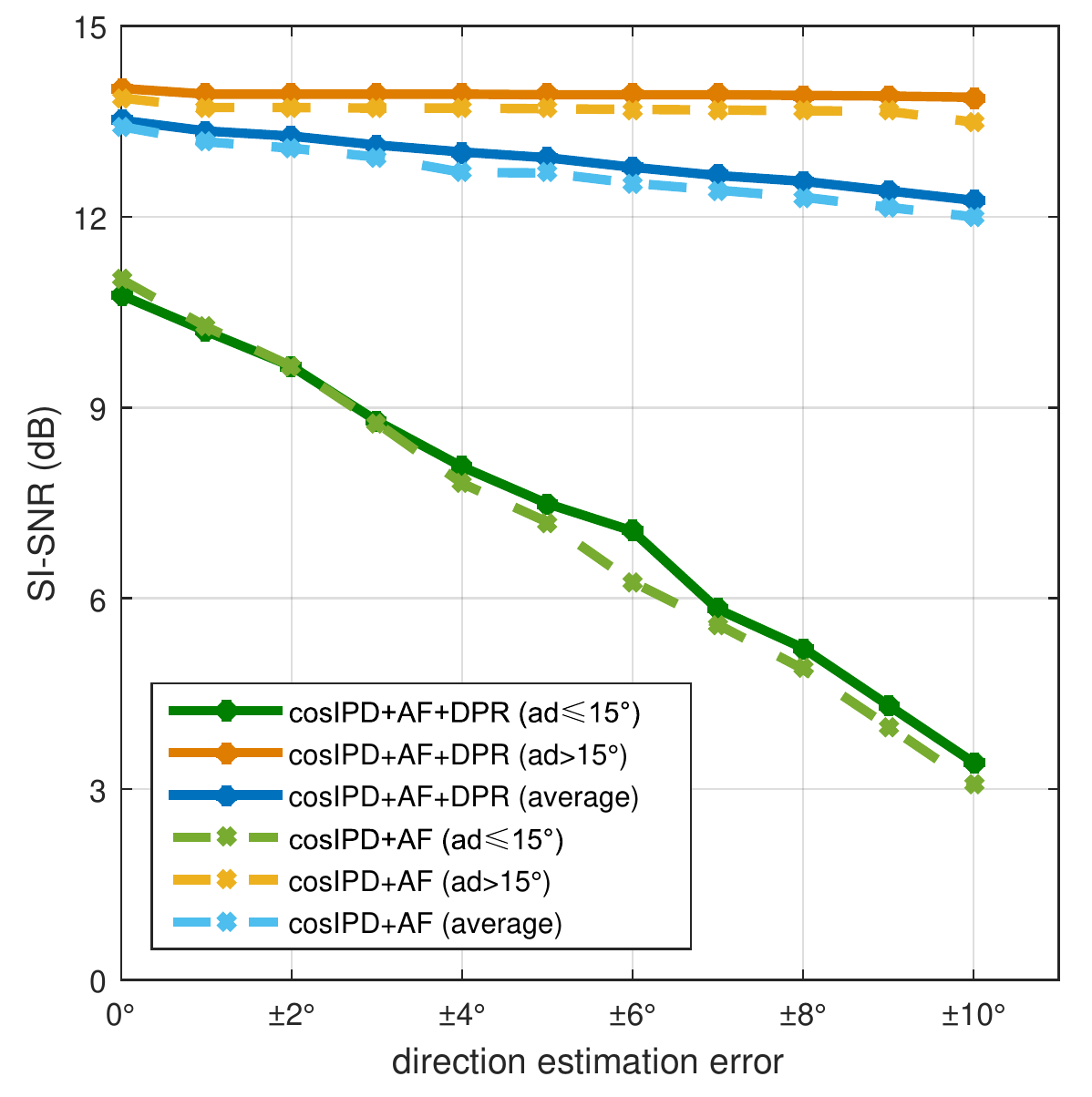}
    \caption{The SI-SDRi performance versus different direction estimation error on spatialized reverberant WSJ0 2-mix.}
    \label{fig:dee_2mix}
\end{figure}

\begin{figure}[h]
    \centering
    \includegraphics[width=7cm]{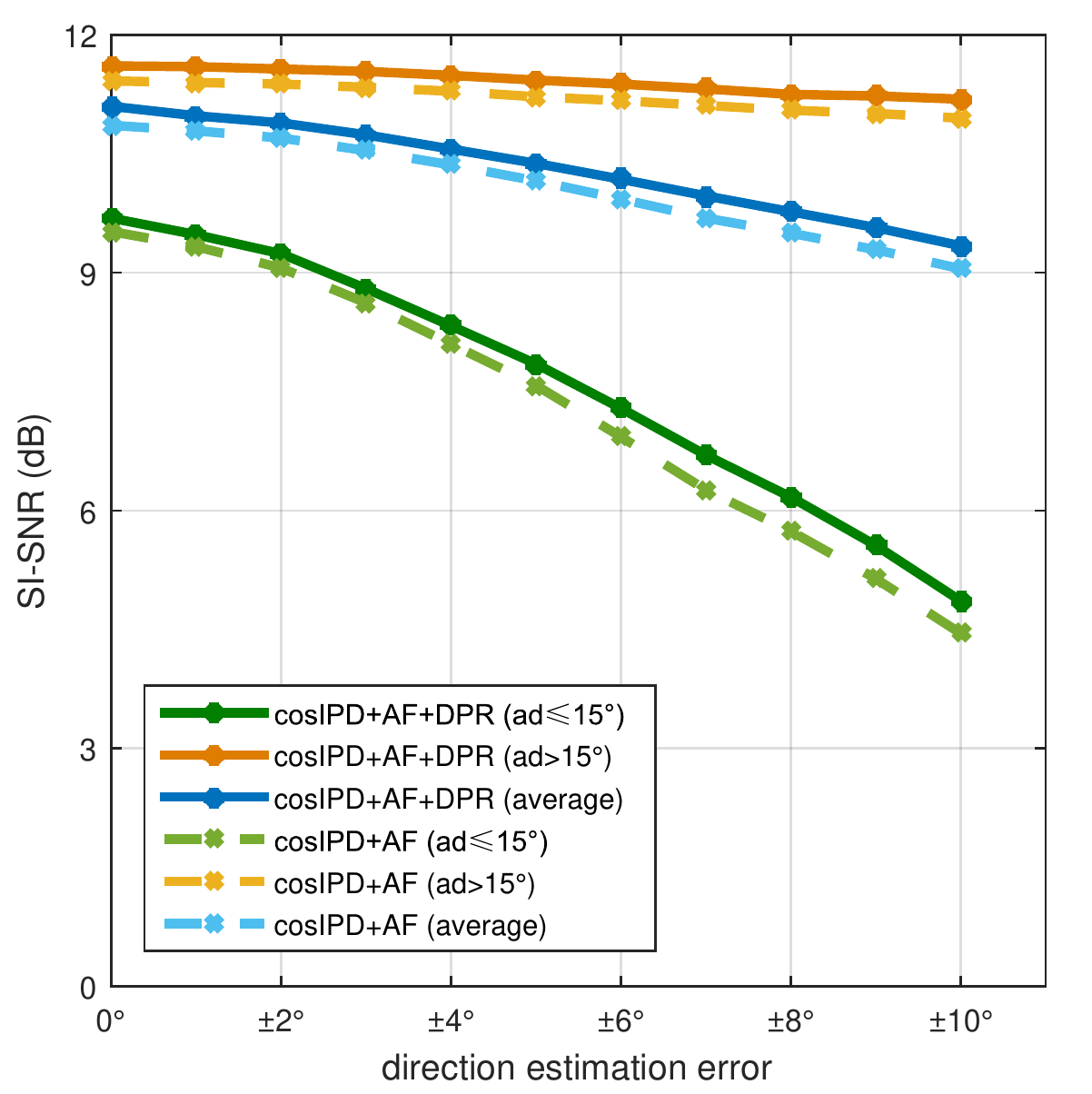}
    \caption{The SI-SDRi performance versus different direction estimation error on spatialized reverberant WSJ0 3-mix.}
    \label{fig:dee_3mix}
\end{figure}

The results for two systems are illustrated in Fig. \ref{fig:dee_2mix} and \ref{fig:dee_3mix}, where the full lines indicate the cosIPD+AF+DPR system while the dotted lines represent the cosIPD+AF system. We can see from the figure that as the direction estimation error increases, the performance of both systems drops drastically when angle difference between 0-15$\degree$. It is reasonable because that, in order to ensure the spatial discrimination within a small angle range, the space grid should be finely divided to distinguish directional sources well. So, even a slight estimation error may lead to a wide deviation to the proper estimation. Also, the estimation error aggravates the spatial ambiguity issue and brings difficulty in determining which speaker should be separated. Encouragingly, figure \ref{fig:dee_2mix} and figure \ref{fig:dee_3mix} tell that when the angle difference is larger than 15$\degree$, the direction estimation error within ±10$\degree$ almost has no influence on the performance of cosIPD+AF+DPR system, i.e., smaller than 0.1dB of SI-SDRi for 2-mix and 0.4dB for 3-mix. While for cosIPD+AF system, the performance drops about 0.4dB for 2-mix and 0.7dB for 3-mix. We achieve this with the aids of DPR feature since it covers a relatively large direction grid (10$\degree$ in our configuration). Also, the overall performance decrease of cosIPD+AF+DPR system is less than 1.2dB on WSJ0-2mix and 1.7dB on WSJ0-3mix. These results confirms the robustness of our proposed spatial filter.

\section{Conclusion}
\label{sec:conclusion}

In this paper, we introduce a temporal-spatial neural filter, which directly estimates the time-domain target speech from multi-speaker mixture in reverberant environments, assisted with  directional information. The main contributions lie in:
1) We propose to jointly model the temporal, spectral, spatial and directional discriminability to create a more complete representation for better separation accuracy;
2) We design two effective directional features to indicate the dominance of directional source at each T-F bin. At meanwhile, experiments demonstrate the robustness of these directional features when existing a direction estimation error.
3) The proposed speaker-independent model can handle mixtures with up to 3 simultaneous speakers and associate the output to the speaker by specifying the speaker's corresponding direction.

Our evaluations showed that the proposed temporal-spatial neural filter outperforms other deep learning based multi-channel speech separation approaches. Also, it has a smaller model size and fast processing speed. Furthermore, we demonstrate that the proposed model is robust to the direction estimation error when speakers are not closely located, i.e., the angle difference between speakers is larger than 15$\degree$.

We point out that this work still has some limitations that will be addressed in future work.
First, one significant flaw of proposed filter is the spatial ambiguity issue. Under small angle different case, the performance is still far from satisfactory. Other target-related information can be introduced to complement the representation, such as speaker embedding \cite{wang2018voicefilter} and visual signals \cite{ephrat2018looking}.
Second, our approach needs to know the direction(s) of speaker(s) in advance, which is a strong assumption. A direction estimation module can be further integrated to build a more complete system.
Third, improved attention mechanism can be introduced to automatically extract the time-varying contribution of each feature. As shown in Fig. \ref{fig:dee_2mix}, when the target direction is precisely estimated, the performance of cosIPD+AF system is about 0.3dB better than that of cosIPD+AF+DPR under small angle difference. This implies that AF alone can provide more separation precision under small angle case. We can further make the model selectively attend to AF and neglect DPR under special scenarios.

Fourth, the proposed method only focuses on speech separation and do not deal with the dereverberation or denoising task. In the future, we will consider training these task jointly.
Finally, we assume that the speaker does not move during speaking and the directional feature is computed with a fixed direction. However, in some practical applications, the directions of speakers may change or jitter during an utterance. For this propose, the directional features should be carefully redesigned and reformulated.


%

\section*{Acknowledgment}
Thanks to Shixiong Zhang, Lianwu Chen, Yong Xu, Meng Yu and Dong Yu from Tencent AI Lab for significant guidance and suggestions. Thanks to Jian Wu from Northwestern Polytechnical University, Xi’an, China, and Fahimeh Bahmaninezhad from University of Texas at Dallas, Richardson, USA for very helpful discussions.

\ifCLASSOPTIONcaptionsoff
  \newpage
\fi



%

\bibliographystyle{IEEEtran}
\bibliography{mybib}

\begin{thebibliography}{10}
\providecommand{\url}[1]{#1}
\csname url@samestyle\endcsname
\providecommand{\newblock}{\relax}
\providecommand{\bibinfo}[2]{#2}
\providecommand{\BIBentrySTDinterwordspacing}{\spaceskip=0pt\relax}
\providecommand{\BIBentryALTinterwordstretchfactor}{4}
\providecommand{\BIBentryALTinterwordspacing}{\spaceskip=\fontdimen2\font plus
\BIBentryALTinterwordstretchfactor\fontdimen3\font minus
  \fontdimen4\font\relax}
\providecommand{\BIBforeignlanguage}[2]{{%
\expandafter\ifx\csname l@#1\endcsname\relax
\typeout{** WARNING: IEEEtran.bst: No hyphenation pattern has been}%
\typeout{** loaded for the language `#1'. Using the pattern for}%
\typeout{** the default language instead.}%
\else
\language=\csname l@#1\endcsname
\fi
#2}}
\providecommand{\BIBdecl}{\relax}
\BIBdecl

\bibitem{Cherry1960Contribution}
C.~Cherry and J.~A. Bowles, ``Contribution to a study of the “cocktail party
  problem”,'' \emph{Journal of the Acoustical Society of America}, vol.~32,
  no.~7, pp. 884--884, 1960.

\bibitem{ephrat2018looking}
A.~Ephrat, I.~Mosseri, O.~Lang, T.~Dekel, K.~Wilson, A.~Hassidim, W.~T.
  Freeman, and M.~Rubinstein, ``Looking to listen at the cocktail party: a
  speaker-independent audio-visual model for speech separation,'' \emph{ACM
  Transactions on Graphics (TOG)}, vol.~37, no.~4, p. 112, 2018.

\bibitem{wang2014training}
Y.~Wang, A.~Narayanan, and D.~Wang, ``On training targets for supervised speech
  separation,'' \emph{IEEE/ACM transactions on audio, speech, and language
  processing}, vol.~22, no.~12, pp. 1849--1858, 2014.

\bibitem{kolbaek2017multitalker}
M.~Kolb{\ae}k, D.~Yu, Z.-H. Tan, J.~Jensen, M.~Kolbaek, D.~Yu, Z.-H. Tan, and
  J.~Jensen, ``Multitalker speech separation with utterance-level permutation
  invariant training of deep recurrent neural networks,'' \emph{IEEE/ACM
  Transactions on Audio, Speech and Language Processing (TASLP)}, vol.~25,
  no.~10, pp. 1901--1913, 2017.

\bibitem{saruwatari2003blind}
H.~Saruwatari, S.~Kurita, K.~Takeda, F.~Itakura, T.~Nishikawa, and K.~Shikano,
  ``Blind source separation combining independent component analysis and
  beamforming,'' \emph{EURASIP Journal on Advances in Signal Processing}, vol.
  2003, no.~11, p. 569270, 2003.

\bibitem{Wang2006CASA}
D.~Wang and G.~J. Brown, \emph{Computational Auditory Scene Analysis:
  Principles, Algorithms, and Applications}.\hskip 1em plus 0.5em minus
  0.4em\relax Wiley-IEEE Press, 2006.

\bibitem{sainath2016factored}
T.~N. Sainath, R.~J. Weiss, K.~W. Wilson, A.~Narayanan, and M.~Bacchiani,
  ``Factored spatial and spectral multichannel raw waveform cldnns,'' in
  \emph{2016 IEEE International Conference on Acoustics, Speech and Signal
  Processing (ICASSP)}.\hskip 1em plus 0.5em minus 0.4em\relax IEEE, 2016, pp.
  5075--5079.

\bibitem{wang2011target}
L.~Wang, H.~Ding, and F.~Yin, ``Target speech extraction in cocktail party by
  combining beamforming and blind source separation.'' \emph{Acoustics
  Australia}, vol.~39, no.~2, 2011.

\bibitem{chen2017cracking}
Z.~Chen, J.~Li, X.~Xiao, T.~Yoshioka, H.~Wang, Z.~Wang, and Y.~Gong, ``Cracking
  the cocktail party problem by multi-beam deep attractor network,'' in
  \emph{2017 IEEE Automatic Speech Recognition and Understanding Workshop
  (ASRU)}.\hskip 1em plus 0.5em minus 0.4em\relax IEEE, 2017, pp. 437--444.

\bibitem{luo2018tasnet}
Y.~Luo and N.~Mesgarani, ``Tasnet: time-domain audio separation network for
  real-time, single-channel speech separation,'' in \emph{2018 IEEE
  International Conference on Acoustics, Speech and Signal Processing
  (ICASSP)}.\hskip 1em plus 0.5em minus 0.4em\relax IEEE, 2018, pp. 696--700.

\bibitem{pfeifenberger2017dnn}
L.~Pfeifenberger, M.~Z{\"o}hrer, and F.~Pernkopf, ``Dnn-based speech mask
  estimation for eigenvector beamforming,'' in \emph{2017 IEEE International
  Conference on Acoustics, Speech and Signal Processing (ICASSP)}.\hskip 1em
  plus 0.5em minus 0.4em\relax IEEE, 2017, pp. 66--70.

\bibitem{zhou2017combined}
H.~Zhou and J.~Lu, ``Combined beamforming and deep neural networks for
  multichannel speech enhancement,'' in \emph{INTER-NOISE and NOISE-CON
  Congress and Conference Proceedings}, vol. 255, no.~4.\hskip 1em plus 0.5em
  minus 0.4em\relax Institute of Noise Control Engineering, 2017, pp.
  3340--3348.

\bibitem{higuchi2016robust}
T.~Higuchi, N.~Ito, T.~Yoshioka, and T.~Nakatani, ``Robust mvdr beamforming
  using time-frequency masks for online/offline asr in noise,'' in \emph{2016
  IEEE International Conference on Acoustics, Speech and Signal Processing
  (ICASSP)}.\hskip 1em plus 0.5em minus 0.4em\relax IEEE, 2016, pp. 5210--5214.

\bibitem{wang2018spatial}
Z.-Q. Wang and D.~Wang, ``On spatial features for supervised speech separation
  and its application to beamforming and robust asr,'' in \emph{2018 IEEE
  International Conference on Acoustics, Speech and Signal Processing
  (ICASSP)}.\hskip 1em plus 0.5em minus 0.4em\relax IEEE, 2018, pp. 5709--5713.

\bibitem{yu2017permutation}
D.~Yu, M.~Kolb{\ae}k, Z.-H. Tan, and J.~Jensen, ``Permutation invariant
  training of deep models for speaker-independent multi-talker speech
  separation,'' in \emph{2017 IEEE International Conference on Acoustics,
  Speech and Signal Processing (ICASSP)}.\hskip 1em plus 0.5em minus
  0.4em\relax IEEE, 2017, pp. 241--245.

\bibitem{wang2019combining}
Z.~Wang and D.~Wang, ``Combining spectral and spatial features for deep
  learning based blind speaker separation,'' \emph{IEEE/ACM Transactions on
  Audio, Speech, and Language Processing}, vol.~27, no.~2, pp. 457--468, 2019.

\bibitem{he2016deep}
K.~He, X.~Zhang, S.~Ren, and J.~Sun, ``Deep residual learning for image
  recognition,'' in \emph{Proceedings of the IEEE conference on computer vision
  and pattern recognition}, 2016, pp. 770--778.

\bibitem{hershey2016deep}
J.~R. Hershey, Z.~Chen, J.~Le~Roux, and S.~Watanabe, ``Deep clustering:
  Discriminative embeddings for segmentation and separation,'' in \emph{2016
  IEEE International Conference on Acoustics, Speech and Signal Processing
  (ICASSP)}.\hskip 1em plus 0.5em minus 0.4em\relax IEEE, 2016, pp. 31--35.

\bibitem{lianwu2019multi}
L.~Chen, M.~Yu, D.~Su, and D.~Yu, ``Multi-band pit and model integration for
  improved multi-channel speech separation,'' in \emph{2019 IEEE International
  Conference on Acoustics, Speech and Signal Processing (ICASSP)}.\hskip 1em
  plus 0.5em minus 0.4em\relax IEEE, 2019.

\bibitem{allen1979image}
J.~B. Allen and D.~A. Berkley, ``Image method for efficiently simulating
  small-room acoustics,'' \emph{The Journal of the Acoustical Society of
  America}, vol.~65, no.~4, pp. 943--950, 1979.

\bibitem{vincent2006performance}
E.~Vincent, R.~Gribonval, and C.~F{\'e}votte, ``Performance measurement in
  blind audio source separation,'' \emph{IEEE transactions on audio, speech,
  and language processing}, vol.~14, no.~4, pp. 1462--1469, 2006.

\bibitem{Benesty2013Study}
J.~Benesty, ``Study and design of differential microphone arrays,''
  \emph{Springer Topics in Signal Processing}, vol.~6, 2013.

\bibitem{chen2017deep}
Z.~Chen, Y.~Luo, and N.~Mesgarani, ``Deep attractor network for
  single-microphone speaker separation,'' in \emph{2017 IEEE International
  Conference on Acoustics, Speech and Signal Processing (ICASSP)}.\hskip 1em
  plus 0.5em minus 0.4em\relax IEEE, 2017, pp. 246--250.

\bibitem{gannot2001signal}
S.~Gannot, D.~Burshtein, and E.~Weinstein, ``Signal enhancement using
  beamforming and nonstationarity with applications to speech,'' \emph{IEEE
  Transactions on Signal Processing}, vol.~49, no.~8, pp. 1614--1626, 2001.

\bibitem{markovich2009multichannel}
S.~Markovich, S.~Gannot, and I.~Cohen, ``Multichannel eigenspace beamforming in
  a reverberant noisy environment with multiple interfering speech signals,''
  \emph{IEEE Transactions on Audio, Speech, and Language Processing}, vol.~17,
  no.~6, pp. 1071--1086, 2009.

\bibitem{xiao2019single}
X.~Xiao, Z.~Chen, T.~Yoshioka, H.~Erdogan, C.~Liu, D.~Dimitriadis, J.~Droppo,
  and Y.~Gong, ``Single-channel speech extraction using speaker inventory and
  attention network,'' in \emph{ICASSP 2019-2019 IEEE International Conference
  on Acoustics, Speech and Signal Processing (ICASSP)}.\hskip 1em plus 0.5em
  minus 0.4em\relax IEEE, 2019, pp. 86--90.

\bibitem{vzmolikova2017learning}
K.~{\v{Z}}mol{\'\i}kov{\'a}, M.~Delcroix, K.~Kinoshita, T.~Higuchi, A.~Ogawa,
  and T.~Nakatani, ``Learning speaker representation for neural network based
  multichannel speaker extraction,'' in \emph{2017 IEEE Automatic Speech
  Recognition and Understanding Workshop (ASRU)}.\hskip 1em plus 0.5em minus
  0.4em\relax IEEE, 2017, pp. 8--15.

\bibitem{luo2019convtasnet}
Y.~{Luo} and N.~{Mesgarani}, ``Conv-tasnet: Surpassing ideal time–frequency
  magnitude masking for speech separation,'' \emph{IEEE/ACM Transactions on
  Audio, Speech, and Language Processing}, vol.~27, no.~8, pp. 1256--1266, Aug
  2019.

\bibitem{wichern2018phase}
G.~Wichern and J.~Le~Roux, ``Phase reconstruction with learned time-frequency
  representations for single-channel speech separation,'' in \emph{2018 16th
  International Workshop on Acoustic Signal Enhancement (IWAENC)}.\hskip 1em
  plus 0.5em minus 0.4em\relax IEEE, 2018, pp. 396--400.

\bibitem{ozerov2010multichannel}
A.~Ozerov and C.~F{\'e}votte, ``Multichannel nonnegative matrix factorization
  in convolutive mixtures for audio source separation,'' \emph{IEEE
  Transactions on Audio, Speech, and Language Processing}, vol.~18, no.~3, pp.
  550--563, 2010.

\bibitem{wang2018deep}
J.~Wang, J.~Chen, D.~Su, L.~Chen, M.~Yu, Y.~Qian, and D.~Yu, ``Deep extractor
  network for target speaker recovery from single channel speech mixtures,''
  \emph{arXiv preprint arXiv:1807.08974}, 2018.

\bibitem{xu2018modeling}
J.~Xu, J.~Shi, G.~Liu, X.~Chen, and B.~Xu, ``Modeling attention and memory for
  auditory selection in a cocktail party environment,'' in \emph{Thirty-Second
  AAAI Conference on Artificial Intelligence}, 2018.

\bibitem{wang2018voicefilter}
Q.~Wang, H.~Muckenhirn, K.~Wilson, P.~Sridhar, Z.~Wu, J.~Hershey, R.~A.
  Saurous, R.~J. Weiss, Y.~Jia, and I.~L. Moreno, ``Voicefilter: Targeted voice
  separation by speaker-conditioned spectrogram masking,'' \emph{arXiv preprint
  arXiv:1810.04826}, 2018.

\bibitem{zmolikova2017speaker}
K.~Zmolikova, M.~Delcroix, K.~Kinoshita, T.~Higuchi, A.~Ogawa, and T.~Nakatani,
  ``Speaker-aware neural network based beamformer for speaker extraction in
  speech mixtures.'' in \emph{Interspeech}, 2017, pp. 2655--2659.

\bibitem{kayser2014estimation}
H.~Kayser, J.~Anem{\"u}ller, and K.~Adilo{\u{g}}lu, ``Estimation of
  inter-channel phase differences using non-negative matrix factorization,'' in
  \emph{2014 IEEE 8th Sensor Array and Multichannel Signal Processing Workshop
  (SAM)}.\hskip 1em plus 0.5em minus 0.4em\relax IEEE, 2014, pp. 77--80.

\bibitem{chen2018multi}
Z.~Chen, X.~Xiao, T.~Yoshioka, H.~Erdogan, J.~Li, and Y.~Gong, ``Multi-channel
  overlapped speech recognition with location guided speech extraction
  network,'' in \emph{2018 IEEE Spoken Language Technology Workshop
  (SLT)}.\hskip 1em plus 0.5em minus 0.4em\relax IEEE, 2018, pp. 558--565.

\bibitem{yoshioka2018multi}
T.~Yoshioka, H.~Erdogan, Z.~Chen, and F.~Alleva, ``Multi-microphone neural
  speech separation for far-field multi-talker speech recognition,'' in
  \emph{2018 IEEE International Conference on Acoustics, Speech and Signal
  Processing (ICASSP)}.\hskip 1em plus 0.5em minus 0.4em\relax IEEE, 2018, pp.
  5739--5743.

\bibitem{chen2018efficient}
Z.~Chen, T.~Yoshioka, X.~Xiao, L.~Li, M.~L. Seltzer, and Y.~Gong, ``Efficient
  integration of fixed beamformers and speech separation networks for
  multi-channel far-field speech separation,'' in \emph{2018 IEEE International
  Conference on Acoustics, Speech and Signal Processing (ICASSP)}.\hskip 1em
  plus 0.5em minus 0.4em\relax IEEE, 2018, pp. 5384--5388.

\bibitem{luo2018surpass}
Y.~Luo and N.~Mesgaran1, ``Tasnet: Surpassing ideal time-frequency masking for
  speech separation,'' \emph{arXiv preprint arXiv:1809.07454}, 2018.

\bibitem{wang2018multi}
Z.-Q. Wang, J.~Le~Roux, and J.~R. Hershey, ``Multi-channel deep clustering:
  Discriminative spectral and spatial embeddings for speaker-independent speech
  separation,'' in \emph{2018 IEEE International Conference on Acoustics,
  Speech and Signal Processing (ICASSP)}.\hskip 1em plus 0.5em minus
  0.4em\relax IEEE, 2018, pp. 1--5.

\bibitem{drude2017tight}
L.~Drude and R.~Haeb-Umbach, ``Tight integration of spatial and spectral
  features for bss with deep clustering embeddings.'' in \emph{Interspeech},
  2017, pp. 2650--2654.

\bibitem{wang2018integrating}
Z.-Q. Wang and D.~Wang, ``Integrating spectral and spatial features for
  multi-channel speaker separation,'' in \emph{Proc. Interspeech}, vol. 2018,
  2018, pp. 2718--2722.

\bibitem{wang2018supervised}
D.~Wang and J.~Chen, ``Supervised speech separation based on deep learning: An
  overview,'' \emph{IEEE/ACM Transactions on Audio, Speech, and Language
  Processing}, vol.~26, no.~10, pp. 1702--1726, 2018.

\bibitem{wang2018alternative}
Z.-Q. Wang, J.~Le~Roux, and J.~R. Hershey, ``Alternative objective functions
  for deep clustering,'' in \emph{2018 IEEE International Conference on
  Acoustics, Speech and Signal Processing (ICASSP)}.\hskip 1em plus 0.5em minus
  0.4em\relax IEEE, 2018, pp. 686--690.

\bibitem{wang2018end}
Z.-Q. Wang, J.~L. Roux, D.~Wang, and J.~R. Hershey, ``End-to-end speech
  separation with unfolded iterative phase reconstruction,'' \emph{arXiv
  preprint arXiv:1804.10204}, 2018.

\bibitem{takahashi2018phasenet}
N.~Takahashi, P.~Agrawal, N.~Goswami, and Y.~Mitsufuji, ``Phasenet: Discretized
  phase modeling with deep neural networks for audio source separation,'' in
  \emph{Proc. Interspeech}, 2018, pp. 2713--2717.

\bibitem{choi2019phase}
H.-S. Choi, J.-H. Kim, J.~Huh, A.~Kim, J.-W. Ha, and K.~Lee, ``Phase-aware
  speech enhancement with deep complex u-net,'' \emph{arXiv preprint
  arXiv:1903.03107}, 2019.

\bibitem{le2019phasebook}
J.~Le~Roux, G.~Wichern, S.~Watanabe, A.~Sarroff, and J.~R. Hershey, ``Phasebook
  and friends: Leveraging discrete representations for source separation,''
  \emph{IEEE Journal of Selected Topics in Signal Processing}, 2019.

\bibitem{mowlaee2012phase}
P.~Mowlaee, R.~Saeidi, and R.~Martin, ``Phase estimation for signal
  reconstruction in single-channel source separation,'' in \emph{Thirteenth
  Annual Conference of the International Speech Communication Association},
  2012.

\bibitem{venkataramani2017adaptive}
S.~Venkataramani, J.~Casebeer, and P.~Smaragdis, ``Adaptive front-ends for
  end-to-end source separation,'' in \emph{Proc. NIPS}, 2017.

\bibitem{lea2017temporal}
C.~Lea, M.~D. Flynn, R.~Vidal, A.~Reiter, and G.~D. Hager, ``Temporal
  convolutional networks for action segmentation and detection,'' in
  \emph{proceedings of the IEEE Conference on Computer Vision and Pattern
  Recognition}, 2017, pp. 156--165.

\bibitem{lea2016temporal}
C.~Lea, R.~Vidal, A.~Reiter, and G.~D. Hager, ``Temporal convolutional
  networks: A unified approach to action segmentation,'' in \emph{European
  Conference on Computer Vision}.\hskip 1em plus 0.5em minus 0.4em\relax
  Springer, 2016, pp. 47--54.

\bibitem{simonyan2014two}
K.~Simonyan and A.~Zisserman, ``Two-stream convolutional networks for action
  recognition in videos,'' in \emph{Advances in neural information processing
  systems}, 2014, pp. 568--576.

\bibitem{feichtenhofer2016convolutional}
C.~Feichtenhofer, A.~Pinz, and A.~Zisserman, ``Convolutional two-stream network
  fusion for video action recognition,'' in \emph{Proceedings of the IEEE
  conference on computer vision and pattern recognition}, 2016, pp. 1933--1941.

\bibitem{wang2018two}
X.~Wang, L.~Gao, P.~Wang, X.~Sun, and X.~Liu, ``Two-stream 3-d convnet fusion
  for action recognition in videos with arbitrary size and length,'' \emph{IEEE
  Transactions on Multimedia}, vol.~20, no.~3, pp. 634--644, 2018.

\bibitem{chollet2017xception}
F.~Chollet, ``Xception: Deep learning with depthwise separable convolutions,''
  in \emph{Proceedings of the IEEE conference on computer vision and pattern
  recognition}, 2017, pp. 1251--1258.

\bibitem{venkataramani2018end}
S.~Venkataramani, J.~Casebeer, and P.~Smaragdis, ``End-to-end source separation
  with adaptive front-ends,'' in \emph{2018 52nd Asilomar Conference on
  Signals, Systems, and Computers}.\hskip 1em plus 0.5em minus 0.4em\relax
  IEEE, 2018, pp. 684--688.

\bibitem{Gannot2017}
S.~Gannot, E.~Vincent, S.~Markovich-Golan, and A.~Ozerov, ``{A Consolidated
  Perspective on Multi-Microphone Speech Enhancement and Source Separation},''
  \emph{IEEE/ACM Transactions on Audio, Speech, and Language Processing},
  vol.~25, pp. 692--730, 2017.

\bibitem{chen2019multi}
L.~Chen, M.~Yu, D.~Su, and D.~Yu, ``Multi-band pit and model integration for
  improved multi-channel speech separation,'' in \emph{ICASSP 2019-2019 IEEE
  International Conference on Acoustics, Speech and Signal Processing
  (ICASSP)}.\hskip 1em plus 0.5em minus 0.4em\relax IEEE, 2019, pp. 705--709.

\bibitem{cardoso1998blind}
J.-F. Cardoso, ``Blind signal separation: statistical principles,''
  \emph{Proceedings of the IEEE}, vol.~86, no.~10, pp. 2009--2025, 1998.

\bibitem{huang2019research}
Y.~Huang, J.~Shi, J.-M. Xu, and B.~Xu, ``Research advances and perspectives on
  the cocktail party problem and related auditory models,'' vol.~45, no.~2,
  2019, pp. 234--251.

\bibitem{naylor2010speech}
P.~A. Naylor and N.~D. Gaubitch, \emph{Speech dereverberation}.\hskip 1em plus
  0.5em minus 0.4em\relax Springer Science \& Business Media, 2010.

\bibitem{pfander2010sparsity}
G.~E. Pfander and H.~Rauhut, ``Sparsity in time-frequency representations,''
  \emph{Journal of Fourier Analysis and Applications}, vol.~16, no.~2, pp.
  233--260, 2010.

\bibitem{colle1976acoustic}
H.~A. Colle and A.~Welsh, ``Acoustic masking in primary memory,'' \emph{Journal
  of verbal learning and verbal behavior}, vol.~15, no.~1, pp. 17--31, 1976.

\bibitem{sawada2010underdetermined}
H.~Sawada, S.~Araki, and S.~Makino, ``Underdetermined convolutive blind source
  separation via frequency bin-wise clustering and permutation alignment,''
  \emph{IEEE Transactions on Audio, Speech, and Language Processing}, vol.~19,
  no.~3, pp. 516--527, 2010.

\bibitem{mandel2017multichannel}
M.~I. Mandel and J.~P. Barker, ``Multichannel spatial clustering using
  model-based source separation,'' in \emph{New Era for Robust Speech
  Recognition}.\hskip 1em plus 0.5em minus 0.4em\relax Springer, 2017, pp.
  51--77.

\bibitem{sawada2006blind}
H.~Sawada, S.~Araki, R.~Mukai, and S.~Makino, ``Blind extraction of dominant
  target sources using ica and time-frequency masking,'' \emph{IEEE
  Transactions on Audio, Speech, and Language Processing}, vol.~14, no.~6, pp.
  2165--2173, 2006.

\bibitem{gannot2017consolidated}
S.~Gannot, E.~Vincent, S.~Markovich-Golan, A.~Ozerov, S.~Gannot, E.~Vincent,
  S.~Markovich-Golan, and A.~Ozerov, ``A consolidated perspective on
  multimicrophone speech enhancement and source separation,'' \emph{IEEE/ACM
  Transactions on Audio, Speech and Language Processing (TASLP)}, vol.~25,
  no.~4, pp. 692--730, 2017.

\bibitem{adel2012beamforming}
H.~Adel, M.~Souad, A.~Alaqeeli, and A.~Hamid, ``Beamforming techniques for
  multichannel audio signal separation,'' \emph{arXiv preprint
  arXiv:1212.6080}, 2012.

\bibitem{li2019tenet}
W.~Li, P.~Zhang, and Y.~Yan, ``Tenet: target speaker extraction network with
  accumulated speaker embedding for automatic speech recognition,''
  \emph{Electronics Letters}, 2019.

\bibitem{le2019sdr}
J.~Le~Roux, S.~Wisdom, H.~Erdogan, and J.~R. Hershey, ``Sdr--half-baked or well
  done?'' in \emph{ICASSP 2019-2019 IEEE International Conference on Acoustics,
  Speech and Signal Processing (ICASSP)}.\hskip 1em plus 0.5em minus
  0.4em\relax IEEE, 2019, pp. 626--630.

\bibitem{vincent2014blind}
E.~Vincent, N.~Bertin, R.~Gribonval, and F.~Bimbot, ``From blind to guided
  audio source separation: How models and side information can improve the
  separation of sound,'' \emph{IEEE Signal Processing Magazine}, vol.~31,
  no.~3, pp. 107--115, 2014.

\bibitem{shi2019furcax}
Z.~Shi, H.~Lin, L.~Liu, R.~Liu, S.~Hayakawa, and J.~Han, ``Furcax: End-to-end
  monaural speech separation based on deep gated (de) convolutional neural
  networks with adversarial example training,'' in \emph{ICASSP 2019-2019 IEEE
  International Conference on Acoustics, Speech and Signal Processing
  (ICASSP)}.\hskip 1em plus 0.5em minus 0.4em\relax IEEE, 2019, pp. 6985--6989.

\bibitem{rao2019target}
W.~Rao, C.~Xu, E.~S. Chng, and H.~Li, ``Target speaker extraction for
  overlapped multi-talker speaker verification,'' \emph{arXiv preprint
  arXiv:1902.02546}, 2019.

\bibitem{gu2019neural}
R.~Gu, L.~Chen, S.-X. Zhang, J.~Zheng, Y.~Xu, M.~Yu, D.~Su, Y.~Zou, and D.~Yu,
  ``Neural spatial filter: Target speaker speech separation assisted with
  directional information,'' in \emph{Interspeech}, 2019.

\end{thebibliography}

%

\end{document}